\documentclass[sigconf]{acmart}

\usepackage{enumitem}
\usepackage{url}

\newcommand{\ignore}[1]{}
\newcommand{\an}[1]{$^{_{_{^{^{#1}}}}}$}

\setcopyright{none}
\renewcommand\footnotetextcopyrightpermission[1]{} 

\begin{document}

\title{The Archives Unleashed Project:\ Technology, Process, and Community to Improve Scholarly Access to Web Archives}

\author{Nick Ruest,\an{1} Jimmy Lin,\an{2} Ian Milligan,\an{3} and Samantha Fritz\an{3}}

\affiliation{\vspace{0.1cm}
$^1$ York University Libraries \\
$^2$ David R. Cheriton School of Computer Science, University of Waterloo\\
$^3$ Department of History, University of Waterloo}

\renewcommand{\shortauthors}{}
\pagestyle{empty} 

\begin{abstract}
The Archives Unleashed project aims to improve scholarly access to web archives through a multi-pronged strategy involving tool creation, process modeling, and community building---all proceeding concurrently in mutually-reinforcing efforts.
As we near the end of our initially-conceived three-year project, we report on our progress and share lessons learned along the way.
The main contribution articulated in this paper is a process model that decomposes scholarly inquiries into four main activities:\ filter, extract, aggregate, and visualize.
Based on the insight that these activities can be disaggregated across time, space, and tools, it is possible to generate ``derivative products'', using our Archives Unleashed Toolkit, that serve as useful starting points for scholarly inquiry.
Scholars can download these products from the Archives Unleashed Cloud and manipulate them just like any other dataset, thus providing access to web archives without requiring any specialized knowledge.
Over the past few years, our platform has processed over a thousand different collections from about two hundred users, totaling over 280 terabytes of web archives.
\end{abstract}

\maketitle

\section{Introduction}

The Archives Unleashed project aims to improve scholarly access to web archives using a three-pronged strategy that simultaneously addresses technology, process, and community.
Our efforts began in 2017 with the generous support of a grant by the Andrew W.\ Mellon Foundation and supplemented by a number of other sources.
As we are nearing the end of the initially-conceived three-year project, the goal of this paper is to share with the broader community experiences we have accumulated along the way.

There is no doubt that scholars are ill-equipped to face the already-here deluge of digital materials for study, but how to provide scholarly access remains a stubborn, unsolved problem.
In this project, our focus is on web archives, but the same can be said of tweets, emails, and a plethora of born-digital records.
As the field of ``web history'' emerges, as defined by the \emph{SAGE Handbook of Web History}~\cite{brugger2018sage} and two recent monographs~\cite{Milligan:2019,Brugger:2018}, this problem is becoming all the more pressing.
Since the rationale for and importance of preserving the web has already been articulated elsewhere, we find no need to repeat those arguments here.

Recognizing that this challenge can only be addressed by a multi-disciplinary team, we assembled one at the outset comprising three main stakeholders:\ librarians and archivists who are charged with gathering and managing digital collections, computer scientists who build the tools necessary to manipulate those collections, and, of course, scholars who interrogate their contents to support various lines of inquiry.
Throughout this paper, we use the term ``scholar'' as a shorthand for the individuals who make use of web archives for their studies, typically historians, digital humanists, or social scientists.
Due to the composition of our team, the needs of historians are perhaps best represented, but our engagements with the broader community through datathons (discussed below) and other workshops have ensured that the social sciences and the digital humanities have not been shortchanged.

The biggest hurdle we set out to overcome is the classic chicken-and-egg problem in technology adoption:\
On the one hand, without guidance from ``users'' (that is, the scholars who study web archives), tool builders are taking stabs in the dark on what the real needs are.
They might produce computational artifacts that are not useful and perhaps tackle the wrong aspects of the overall challenge.
On the other hand, without any existing tools as reference points and often lacking the technical training to understand ``what's possible'', scholars may have a hard time articulating their needs.
Without a community of eager scholars, tool builders are often hesitant to invest in software development efforts that might not lead to meaningful adoption.
And of course, without the right computational tools, scholars cannot make headway in their inquiries.
Thus, we were at an impasse.

The goal of this paper is to report how we have made progress in providing scholarly access to web archives through a multi-pronged strategy involving tool creation, process modeling, and community building---all proceeding concurrently in mutually-reinforcing efforts.
Although the problem remains far from solved, significant progress has been made:
Our contributions comprise lessons learned that are hopefully valuable to the rest of the community and can be transferred beyond web archives into other domains of digital preservation as well.

\section{Project History}
\label{section:history}


Warcbase~\cite{Lin_etal_JOCCH2017}, the immediate predecessor to our current project, was conceived as a scalable web archiving platform to support temporal browsing and large-scale analytics.
Development of the platform began in 2013 by one of the co-authors of this paper (Lin), a computer scientist working on big data, and lacking input from scholars, it suffered exactly from the ``tool builder without users'' problem discussed above.
At around the same time, another one of the co-authors (Milligan), a historian studying early web history, faced the exact opposite problem:\ he was engaging in scholarly inquiry at scale without the proper computational tools and thus struggled to make headway.
A fortuitous and timely gathering of the co-authors, along with many like-minded individuals at the Working with Internet Archives for Research (WIRE) Workshop in 2014 planted the seeds of what would eventually evolve into the Archives Unleashed project.

In 2015, a group of colleagues who had coalesced at the workshop (including the two co-authors mentioned above) collectively stumbled upon a potential solution to the scholarly access problem:\ to overcome the chicken-and-egg problem by building tools {\it and} community {\it simultaneously}.
Through a series of in-person ``datathons'' dubbed ``Archives Unleashed'', the group brought together a few dozen stakeholders to engage in community building and skills training~\cite{Milligan_etal_JCDL2019}.
Over the course of two to three days, tool builders (typically, computer scientists) interacted with scholars, librarians, and archivists with the goal of overcoming the challenges outlined above.
The first datathon was held in Toronto in March 2016, primarily with the support of the Social Sciences and Humanities Research Council of Canada and the U.S.\ National Science Foundation (plus additional contributions from a host of other organizations).
The event successfully brought together a group of stakeholders to initiate the community building process, which continued through three more events, in June 2016, February 2017, and June 2017.

The present Archives Unleashed project retained the catchy moniker from these previous datathons and officially began in June 2017 with a different team.
By that time, the approach to simultaneously building technology and community seemed to be gaining traction, and thus it made sense to build on those initial successes.
We additionally recognized the importance of {\it process} (more details below), thus leading to a three-pronged strategy:

\begin{enumerate}[leftmargin=*]

\item {\it Development of the Archives Unleashed Toolkit (AUT).}
This open-source toolkit represents the evolution of the Warcbase platform, with the benefit of a better understanding of scholars' needs, gained through both the composition of our team and experiences from the datathons.

\item {\it Deployment of the Archives Unleashed Cloud (AUK).}
To bridge the gap between open-source tools and their deployment at scale, we built a one-stop portal that allows users to ingest their collections and execute a number of analyses with a few keystrokes and mouse clicks.
This service operationalizes a process model for scholarly inquiry that provides guidance on how to interrogate collections at scale.
Although the Archives Unleashed Toolkit is open source, which means that anyone is able to install and run the software, we expect that most scholars would be uninterested in managing their own infrastructure.
Thus, AUK can be viewed as the ``canonical instance'' of AUT where we handle deployment and maintenance, freeing scholars to focus on their inquiries.

\item {\it Organization of Archives Unleashed Datathons.}
These efforts at community-building and outreach represent a continuation of the previous in-person events, but with a shift in focus to the tools and services that the project was developing.
In addition, we used these venues as an opportunity to develop longer-term plans to {\it sustain} the scholarly community around web archiving beyond the life of the project.

\end{enumerate}

\noindent In order to more easily make headway in addressing these goals, our efforts primarily focused on thematic web archive collections being gathered by a diverse group of small to medium cultural heritage institutions, specifically those who are subscribers to Internet Archive's Archive-It service.
These subscribers represent the vast majority of institutions in the United States that engage in active web archiving efforts; the 2017 National Digital Stewardship Alliance Web Archiving Survey Report found that 94\% of institutions were using Archive-It that year, with an additional 4\% collecting via separate Internet Archive contracts~\cite{ndsa2017}.
Individual collections of these sort typically range from tens of gigabytes to (a small number of) terabytes.
Quite specifically, collection development and content harvesting lie outside the scope of our project, as there are already many existing resources that provide guidance on those aspects of the web archiving lifecycle.
Thus, we assume that an organization already has a collection of WARC files (the standard container format for web archives) or ARC files (an older file format), in most cases already held by the Internet Archive as part of Archive-It, and desires to provide scholarly access to them.

We believe that targeting Archive-It subscribers maximizes our potential impact.
By definition, these organizations already recognize the need for web archiving, and most of them have already begun to harvest content that meets their institutions' development needs.
These small-to-medium organizations often operate with barebones staffing, and hence are not in a position to actively facilitate scholarly access---yet they truly feel the pain of letting their collections lie fallow.
Our project aims to answer the question: ``We've begun a web archiving program and have gathered a few collections, now what?''
While our toolkit may also be useful to large organizations, for example, national libraries, they are not the primary audience we intend to serve.

Given this brief overview of the genesis of the Archives Unleashed project, the remainder of this paper will focus on lessons we have learned along the way that we hope will be valuable to the broader community.

\section{High-Level Lessons}

As with most technology adoption challenges, perhaps unsurprisingly, the technology itself (AUT and AUK in our case) is relatively straightforward.
The toolkit and the cloud service represent fairly standard instances of open-source software development, and we have adopted standard best practices that lead us down relatively well-trodden paths.
From the technical perspective, the biggest innovation in the toolkit (beyond its Warcbase foundations) is the move from Spark's resilient distributed datasets (RDDs) to DataFrames and a shift from Scala to Python as the language of choice (see Section~\ref{section:toolkit} for more details).

We have come to realize that the {\it process} of scholarly inquiry (in the face of the daunting sizes of many collections) is the most critical component in building and sustaining a community around web archiving.
This is not to diminish the importance of getting the community together in the first place (the third prong of our strategy), a challenge our datathons have already made some headway on.
However, even once we get all the stakeholders ``in the same room'', they still need to ``do something'' to engage in scholarly inquiry, and that ``something'' can leverage our proposed process model to serve as a framework for organizing their activities.
We had previously proposed what we called the FAAV cycle~\cite{Lin_etal_JOCCH2017} to characterize how scholars interrogated web archives, comprised of four main activities:\ filter, analyze, aggregate, and visualize.

In this paper, we present a refinement to FAAV that we term FEAV, where ``analyze'' has been replaced with ``extract'' for a more accurate characterization.
This process model provides a descriptive characterization of scholarly activity based on our observations and can be used prescriptively in a pedagogical manner.
It offers an organizing framework for the rest of the project, guiding the development of the toolkit as well as the cloud service.

The articulation of the FEAV process represents a major contribution of our work.
Our single most important insight is that the main activities (filter, extract, aggregate, visualize) can be disaggregated across time, space, and tools.
That is, each of the activities need not occur in the same session or even the same location, and perhaps most importantly, with the same tools.
More concretely, we have realized that a number of ``derivative products'' provide useful starting points to scholarly inquiry, more so than the raw collections themselves (at least at the outset).
These derivative are essentially the output of a pre-determined pipeline of filtering, extraction, and aggregation that we explicitly store and share.
In addition to being useful from a scholarly perspective, these derivatives are also much smaller than the raw collections, typically manipulable on a laptop or in a cloud notebook.
These derivative and the disaggregation of FEAV yield important implications for adoption, which we detail in Section~\ref{section:derivatives}.

\section{The FEAV Process Model}
\label{section:process}

As we have argued above regarding the central role of {\it process} in facilitating scholarly access to web archives, it makes sense to begin with a detailed discussion of our FEAV model.

Our previous work has considerably influenced this process model.
Even prior to the Archives Unleashed project, in the context of Warcbase, we aimed to serve a well-defined group of scholars as our primary audience.
Although we did not expect these scholars to have formal computer science training, our initial efforts targeted those who were already comfortable with a scripting language such as Python or R.
We assumed that the scholars could perform simple manipulations of datasets in well-defined formats (e.g., CSV or JSON), use standard libraries to execute common tasks (e.g., create a word cloud), or were proficient enough with search engines to figure out what to do by searching online for instructions.
Furthermore, we expected that the scholars were already comfortable with the command line and possessed basic knowledge of the file system (e.g., moving or copying files and directories).

In our experience, such a skillset is relatively common, particularly among recently-trained scholars who want to seriously work with digital objects.
It might not be unrealistic to expect that these form the ``core compentencies'' of {\it all} future scholars in this intellectual space, but how to bring everyone up to this level of technical proficiency is beyond the scope of our work.
More importantly, even if graduate schools and professional organizations within the humanities and social sciences are not properly training scholars in these skills, they are competencies obtainable through libraries and other free resources.
The Programming Historian, for example, has existed in various formats since 2008~\cite{maria_jose_afanador_llach_2019_3525082} and Software Carpentry (and related projects such as Data Carpentry) has been running in-person workshops around the world on basic computational skills and principles.
A growing body of literature explores best practices for how best to develop collaborative lessons for these diverse learners~\cite{10.1371/journal.pcbi.1005963}.
In other words, it would truly be a bridge too far in most cases for a historian to learn how to parse one hundred gigabytes of raw WARC files without assistance; but, with the support of appropriate training and resources (such as those referenced above), scholars can be reasonably expected to develop fluency in using Jupyter notebooks to manipulate modest amounts of CSV or JSON data (for example).

\subsection{From Madlibs to a Process Model}

For our target group of scholars, early datathons showed that they generally had little difficulty grasping the concepts (e.g., transformations on large collections of data records, accessing fields in a tuple, etc.)\ and the ``mechanics'' of using Warcbase (and later, the Archives Unleashed Toolkit).
However, when it became time to analyze their own collections, the scholars were often unsure where to begin~\cite{Lin_etal_JOCCH2017}.
Indeed, we were faced with this conundrum:\ with us in the room to provide support, scholars could perform amazingly creative analyses, but few adopted the technology on their own without assistance.
It soon became clear that scholars did not know where to start.
Faced with several hundred gigabytes of WARC files and a command shell (with a blinking cursor) ready to accept commands, what to do?

We have attempted to jumpstart the process of inquiry by providing a number of examples illustrating the capabilities of our tools, adopting a ``madlibs'' (i.e., fill-in-the-blank) approach; more in Section~\ref{section:notebooks}.
For example, we illustrate:\ this is how you find the website that has the most mentions of Canadian Prime Minister Justin Trudeau---if you want to change the person of interest and the collection, change the variables {\it here} and {\it there}.
Or, this is how you create a word cloud of the crawl from September 2014---you can change the temporal interval using {\it this} variable, or focus on a particular domain by setting {\it that} variable.

What began to emerge, essentially, was a ``cookbook'' with a series of ``recipes'' for addressing a number of common analytics tasks.
Indeed, this format now forms the backbone of our online documentation.\footnote{\url{https://github.com/archivesunleashed/aut-docs}}
These recipes can be phrased in the form of ``how do I...'' questions, for example:

\begin{itemize}[leftmargin=*]
\item extract all URLs in a collection?
\item count the occurrences of different domains?
\item extract plain text from URLs matching a pattern?
\item compute counts of specific keywords?
\item find the most common person mentioned?
\item extract the anchor text of links to a particular URL?
\item find all pages that link to a site?
\item determine the most popular gif?
\item compute the checksum of images?
\end{itemize}

\noindent With this cookbook approach, combined with observations of how our tool was being used successfully in the datathons and through interactions with other scholars, we began to generalize scholarly activities into a process model we termed the FAAV cycle, first articulated in Lin et al.~\cite{Lin_etal_JOCCH2017}.
FAAV stands for ``filter'', ``analyze'', ``aggregate'', and ``visualize''---the four main activities we found scholars engaging in as part of their inquiries.
Over the past few years, we have further refined FAAV into FEAV, which we describe in detail next:\ the main refinement is replacing ``analyze'' with ``extract'' for greater descriptive accuracy.

\subsection{The Four Main Activities}
\label{section:4activities}

The four main activities of the FEAV model are ``filter'', ``extract'', ``aggregate'', and ``visualize'' in support of scholarly inquiry:

\smallskip \noindent {\bf Filter.}
Typical web archive collections range from tens of gigabytes to terabytes; it is relatively rare---with perhaps the exception of high-level exploratory probes and ``distant reading'' studies from the greatest distance---that an individual analysis would consume the {\it entire} collection.
Thus, a scholar usually begins by focusing on a particular subset of the web archive, which we characterize as filtering.
This can be accomplished by content, metadata, or some extracted information.
Content-based filtering might be based on text, e.g., consider only pages mentioning a particular set of keywords, or based on the hyperlink structure between the pages, e.g., consider only pages that link to a particular website or domain.
Metadata-based filtering might be based on a particular range of crawl dates or pages whose URLs match a particular pattern.
Filtering can also be based on any information extracted from either the content or metadata, as the result of running arbitrary user-defined functions (more below).
The filtering criteria may be arbitrarily complex and nested, for example, a scholar is only interested in pages containing a particular keyword that link to a specific domain, within some temporal range.

\smallskip \noindent {\bf Extract.}
After selecting a subset of material, the scholar typically then extracts some information of interest.
Examples include extracting the plain text from the raw HTML source, identifying mentions of named entities (e.g., people, organizations, places), assessing the sentiment of the underlying text, etc.
Extractions need not be limited to HTML---for example, a scholar might be interested in PDFs in a collection---or even be limited to textual data---we have begun experimenting with pipelines that analyze images, for example.
Typically, extraction is accomplished by user-defined functions (UDFs); the user in this case refers to the programmer.
Generally, we would not expect scholars to write their own UDFs (at least at the outset); instead, we provide a library of UDFs to perform common operations that they can draw from and assemble in novel combinations.
Note that UDFs are extensible and can invoke arbitrarily-complex functionalities---for example, leveraging a deep neural network to perform object detection on images~\cite{yang2019}.

\smallskip \noindent {\bf Aggregate.}
The output of filtering and extraction is a collection of records of interest, typically already far smaller (often, orders of magnitude smaller) than the raw collection.
In most cases, however, these records need to be aggregated or summarized before they are suitable for human consumption.
The simplest example of aggregation is to produce a table of counts, e.g., how many times a person or location has been mentioned within a set of pages, how many times a particular page has been linked to, etc.
Other common examples include finding the maximum (e.g., the page with the most incoming links), the minimum (e.g., the least frequently-mentioned name from some list of individuals), or the average (e.g., the average of sentiment expressed across pages of a website as determined by an automatic classifier).

\smallskip \noindent {\bf Visualize.} 
Finally, the aggregated results are presented in some sort of visualization for the scholar's consumption.
The visualization can be as simple as a table or list showing individual records, directly generated by our toolkit, or the output of the toolkit can be passed to an external application to generate complex, interactive visualizations.
These visualizations can be static (e.g., a figure, graph, or plot) or interactive;
they can be designed to support further exploration (i.e., an intermediate product) or be prepared for a publication or a blog post (i.e., for dissemination).
It is not our intention for the Archives Unleashed Toolkit to evolve into a comprehensive visualization framework; instead, our main goal is to support interoperability with existing visualization toolkits in the broader ecosystem that scholars may already be familiar with.

\subsection{Discussion}

Although we describe the FEAV process model as comprised of distinct activities, it is important to note that all of the activities are closely intertwined in practice, and may not even proceed in the described order.
For example:

Does a scholar filter first, then perform extraction, or vice versa?
Sometimes, filtering is performed on extraction results---for example, running a named-entity detector to identify person names, and then considering only pages that mention certain names.
In other cases, the distinction between filtering and extraction is blurred---for example, filtering based on hyperlinks technically requires running a link extractor on the page first, but conceptually, the scholar's goal is to filter, not to extract.
In practice, filtering and extraction are often tightly coupled.

Filtering is also commonly applied to the output of aggregations.
After counting, the scholar might wish to discard items that appear too frequently or too rarely.
In web collections, there is inevitably a long tail of items that appear only a small number of times (e.g., misspelled names), and for the most part, scholars are not interested in those.
On the flip side, there are often items that appear very frequently (e.g., ``here'' as anchor text) and hence it makes sense to discard those as to not clutter up the analysis.
The combination of these activities in our process model, we argue, is flexible enough to accommodate diverse scholarly needs.

Finally, filtering, extraction, and aggregation might proceed cohesively in a tight loop:
For example, an initial set of filtering keywords yields too many pages and thus requires additional refinement to produce a manageable set of results.
Or the opposite could happen---the filter was too restrictive and did not retain enough records for analysis.
These activities might even alter the course of scholarly inquiry---for example, a result leads to an interesting question that compels the scholar to follow another tangential thread.

Note that during the normal course of scholarly inquiry, some of the activities might be skipped altogether.
For example, if the filter is very specific and retains only a handful of records (e.g., half a dozen pages), then the scholar may decide to examine those results directly---in this case, there is no meaningful aggregation and visualization to speak of.
By varying the specificity of the filter, a scholar can switch between  ``distant'' or ``close'' reading~\cite{Moretti_2007} when using our toolkit.
Once again, these possibilities are captured by our process model.

\subsection{Standard Derivatives}
\label{section:derivatives}

Both our original FAAV and the refined FEAV model emerged as a descriptive, bottom-up characterization of what we observed scholars doing as they engaged with web archives.
It was not our original intent that the model be applied prescriptively---our goal was merely to provide a reference that scholars could consult.
Later on, though, we realized that our model could indeed serve a pedagogical purpose; that is, during tutorial sessions in our datathons, we offered FEAV as a way to help new users of the toolkit structure their approach to inquiry.
We would offer:\ once you've formed your initial research question, think about what subset of the collection you would need to interrogate, what information you'd need to extract from those pages, etc.

As a result of this guidance, many analyses began much the same way, with very similar initial steps.
In particular, we have found three analysis products to be so frequently used by scholars that with the Archives Unleashed Cloud and for collections used at our datathons, we have begun to pre-generate and store them, before they are even requested.
Among other benefits, this obviates the need for scholars to run potentially large analysis jobs before they arrive at a datathon, saving precious time in our in-person events.
These are what we term our three ``standard derivatives'', the creation of which is tightly integrated into the Archives Unleashed Cloud (see Section~\ref{section:cloud} for more details):

\begin{itemize}[leftmargin=*]

\item {\bf Domain Distribution.} We extract all URLs to compute the frequency of domains appearing in the collection.

\item {\bf Domain webgraph.} We extract all hyperlinks to create a domain-to-domain network graph; that is, the hyperlink structure of the collection, aggregated at the domain level.

\item {\bf Plain Text.} We extract plain text from all web pages, along with metadata such as crawl date, domain name, and the URL. 

\end{itemize}

\noindent It remains an interesting question as to whether these derivatives are popular because we provide them, or if their popularity is in response to scholarly demand.
The answer is likely a mixture of both, but these derivatives have two important properties:\
First, they are genuinely useful in answering a number of basic questions that are applicable to {\it every} web archive:
The domain distribution provides a starting point to answering the question ``What's in this collection?''
It also provides the scholar with a sense of potential biases that may be present---by noting, for example, the over- or under-representation of certain sites.
The domain webgraph provides a tractable overview of the collection structure---after all, the hyperlink structure is one important distinguishing characteristic that separates web archives from other large collections of documents.
Page-to-page link structure is usually too fine-grained to be helpful in providing an overview, and we have found that domain-based aggregation provides a nice middle ground that shows interesting structures without being overwhelming in size.
The plain text of pages is useful for obvious reasons, serving as the starting point for most content-based analyses.

Second, and perhaps more importantly, these derivatives are often small enough to be manipulable on the scholar's laptop.
Our previous work~\cite{Deschamps_etal_JCDL2019_costs} provides a detailed quantitative analysis of the relationship between raw and derivative sizes, but in rough terms, for a ``typical'' Archive-It collection, the domain distribution data is usually less than 1 MB, the domain webgraph is 10s MB, and the raw text is perhaps 10s GB.

An apt analogy to the creation and use of these derivatives might be the notion of an opening game in chess, which formulaically captures combinations of the first few moves in chess games.
The existence of fixed opening moves and well-known countermoves certainly does not diminish the overall beauty of the game or the creativity necessary to play---in the same way that our derivatives do not diminish the creative potential of scholarly inquiry.
These products merely present useful starting points, and scholars always have the option of starting from the raw web archives.

\subsection{FEAV Disaggregation}
\label{section:disaggregation}

\begin{figure}
    \centering
    \includegraphics[width=0.7\columnwidth]{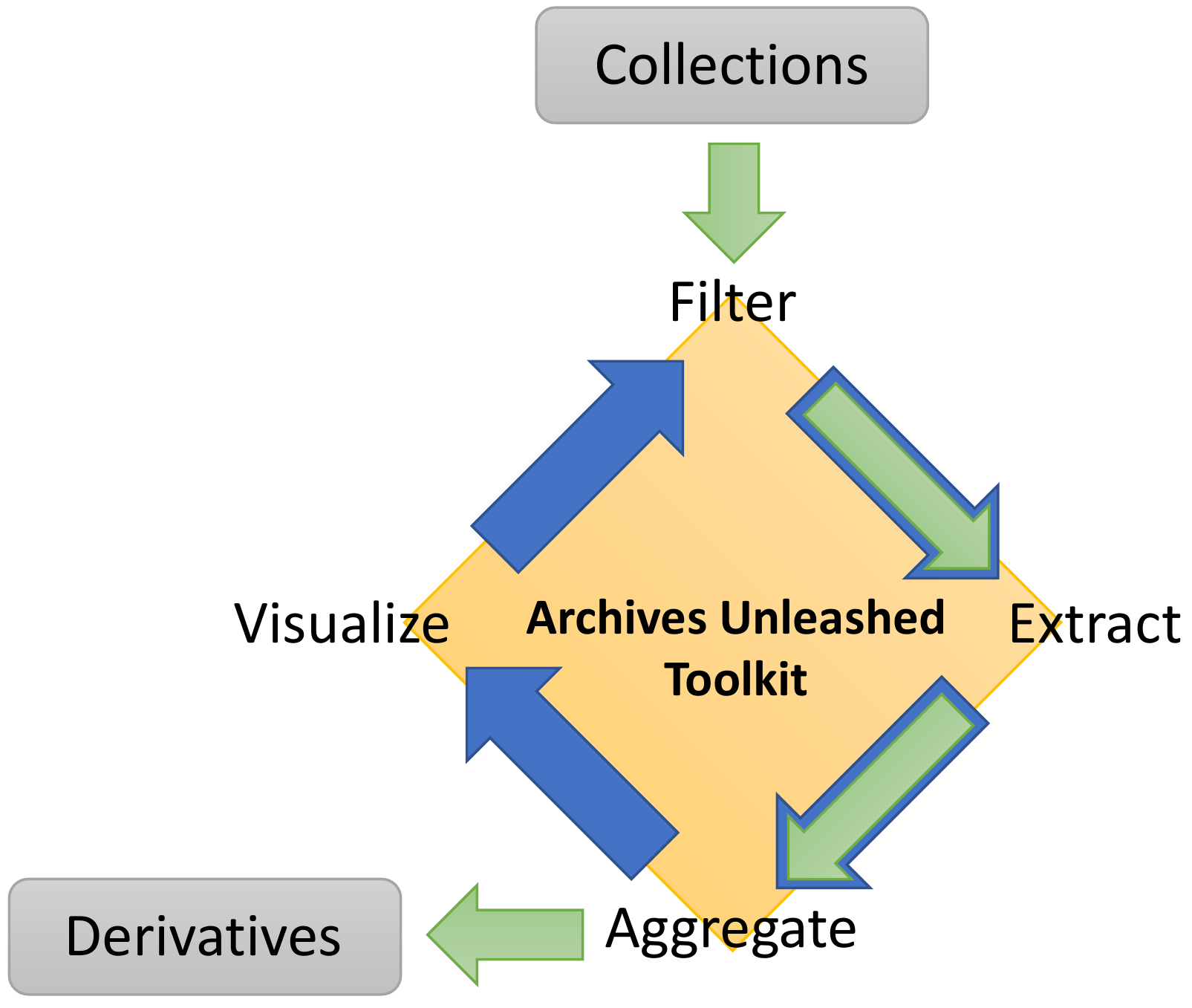}
    \vspace{-0.1cm}
    \caption{The relationship between our FEAV process model and the Archives Unleashed Toolkit and Cloud.
    The toolkit is designed to support all four main activities, shown by the blue arrows.
    The cloud, in contrast, is designed to only support a subset of the activities, but additionally allows users to manage the ingestion of collections and downloading of derivatives (green arrows).}
    \label{fig:cycle}
    \vspace{-0.15cm}
\end{figure}

\begin{figure*}[t]
\centering
\begin{minipage}[t]{.49\textwidth}
\begin{center}{\small
\begin{verbatim}
RecordLoader.loadArchives("collection/", sc)
 .keepValidPages()
 .flatMap(r => ExtractLinks(r.getUrl, r.getContent))
 .map(r => (ExtractDomain(r._1), ExtractDomain(r._2)))
 .filter(r => r._1 != "" && r._2 != "")
 .countItems()
 .filter(r => r._2 > 5)
 .saveAsTextFile("domain-webgraph/")
\end{verbatim}}
\end{center}
\end{minipage}
\begin{minipage}[t]{.49\textwidth}
\begin{center}{\small
\begin{verbatim}
RecordLoader.loadArchives("collection/", sc)
  .webgraph()
  .groupBy(ExtractDomain($"src"), ExtractDomain($"dest"))
  .count()
  .filter($"count" > 5)
  .write.csv("domain-webgraph/")
\end{verbatim}}
\end{center}
\end{minipage}
\caption{Simplified Scala scripts for generating the domain webgraph using the Archives Unleashed Toolkit.
On the left we present the analysis using resilient distributed datasets (RDDs); on the right we present the same analysis using DataFrames.}
\label{fig:aut-example-code}
\vspace{-0.1cm}
\end{figure*}

Somewhat surprisingly, we have found the standard derivatives described above to be so useful that we encourage scholars (particularly those just learning about web archives) to use them as starting points of analyses, rather than the raw archives themselves.
This has led to the single biggest insight that has facilitate adoption---the recognition that FEAV can be disaggregated across time, space, as well as tools.
What does this mean?
Allow us to explain:

The derivatives represent a combination of filtering, extraction, and aggregation whose output we capture, store, and share.
Thus, when using these products, the scholars are taking the output of FEA and continuing with their own iteration of FEAV.
At that point, how these derivatives came to be generated (when and where) becomes irrelevant (this is exactly the disaggregation we speak of), since scholars can engage with the material on their own laptops (a different time, a different place).
Furthermore, scholars are not limited to analyses with our toolkit---they can use whatever tools they are already familiar with, be it Python, R, and yes, even Microsoft Excel.
In this way, our process model supports seamless integration of our tools and services with the broader ecosystem.
Given the popularity of Python for data science today, scholars can conduct analyses in notebooks, manipulate the derivatives using Pandas DataFrames, use existing packages for visualization, or take advantage of a multitude of other capabilities in the Python ecosystem.
This means that scholars can begin inquiries into web collections without needing to know anything specific about web archives (for example, the difference between an ARC and a WARC or the subtleties in parsing hyperlinks to recover anchor texts).

These relationships are shown in Figure~\ref{fig:cycle}.
With the Archives Unleashed Toolkit, a scholar can directly engage in the FEAV processes (with perhaps the support of external applications for visualization)---these are indicated by the blue arrows.
This, of course, requires the scholar to download, install, and configure the toolkit, presenting a potential barrier to entry, but offering the most complete suite of capabilities.
Alternatively, the scholar can directly download the derivatives from the Archives Unleashed Cloud.
Behind the scenes, the portal is still using the toolkit to perform filtering, extraction, and aggregation, but these steps are hidden from the scholar.

It is somewhat ironic that the success of our project means making our project invisible to scholars.
For example, a social scientist downloads the domain webgraph of a particular collection, loads the structure into a Pandas DataFrame in Python, and proceeds to analyze the clustering properties of a number of websites of interest.
She is unfamiliar with how the webgraph was extracted (through an AUT job triggered by AUK in the cloud, see Section~\ref{section:cloud}), but those details are unimportant; the upshot is that we have enabled her research in an unobtrusive manner.
The scholar already knew how to manipulate DataFrames using Pandas, and so analyzing web archives did not require learning any new skills.

We have fully embraced this philosophy of making our tools and services as invisible as possible, because the best strategy to adoption is to not require users to learn anything new!
In this way, the Archives Unleashed Cloud has evolved into a conduit that allows users to ingest raw web archive collections and produce these derivatives for easy access.

\section{The Toolkit}
\label{section:toolkit}

The Archives Unleashed Toolkit (AUT) is the successor to Warcbase and can be characterized as a refinement of the older package, as opposed to an entirely new platform.
For this reason, we focus on differences and new features, and avoid duplicating previously-published material; see Lin et al.~\cite{Lin_etal_JOCCH2017} for additional details.

At a high level, AUT provides a domain-specific language for analyzing web archives built on top of
the open-source Apache Spark data processing platform.
Previously, Warcbase also supported temporal browsing via HBase, a distributed, scalable, big data store that supports low-latency random access to large datasets.
This feature, however, was little used, since scholars were already accustomed to using the Internet Archive's Wayback Machine for temporal browsing; hence, AUT removed such support from Warcbase.

On top of Spark's core data structure, known as resilient distributed datasets (RDDs), the Archives Unleashed Toolkit provides three main capabilities specific to web archiving:

\begin{itemize}[leftmargin=*]

\item {\it Input and output connectors.}
Our toolkit provides abstractions to handle robust parsing of WARC and ARC container files that comprise standard web archive collections, exposing RDDs with records containing usable fields such as HTML content and HTTP headers (for web pages).
It is easy to understate the amount of effort that has been devoted to building robust input connectors: throughout the project we have encountered countless errors, corner cases, and non-compliant data across hundreds of terabytes of web archives, each of which we've had to manually debug and build error handling code for.
This has likely been the biggest sink of development effort.
In addition to input connectors, the toolkit also includes output connectors to facilitate interoperability with external applications, for example, exporting webgraphs in the GraphML format to be used with the Gephi graph visualization platform.

\item {\it Library of user-defined functions (UDFs) for extraction.}
To support extraction of useful information from raw individual records, the toolkit provides a collection of UDFs to conduct various analyses.
These include functions for manipulating URLs, working with text, hyperlinks, images, etc., and more.
This library is constantly evolving in response to scholars' needs.

\item {\it Convenience transformations.}
Spark programs are typically described as a sequence of transformations operating on RDDs.
Many commonly-used sequences of transformations, such as filtering RDDs to retain records of a certain type (e.g., HTML pages) as well as a number of frequently-used aggregations, are encapsulated in ``convenience transformations'' to reduce the verbosity of analysis scripts.

\end{itemize}

\noindent A simplified AUT script (in Scala) for generating the domain webgraph (i.e., one of our standard derivatives, see Section~\ref{section:derivatives}) is shown on the left side of Figure~\ref{fig:aut-example-code}.
The script has been simplified for presentation purposes, but retains the same conceptual structure as the actual working version.
It begins by invoking the \texttt{RecordLoader} to load a collection:\ note here that we hide details such as compression, processing of multiple files, WARC vs.\ ARC formats, etc.
The scholar only needs to specify the path of the directory containing the data.
Next, \texttt{keepValidPages} is a convenience transformation that filters the raw archive RDD to retain only valid HTML pages.
The transformation hides many details that go into the decision of whether a record is a ``valid'' HTML page, based on the server's response MIME type, file extension, and other details.

After the collection has been filtered to retain only the HTML pages, hyperlinks are extracted, and from the source and target of the hyperlinks we keep only the domain (the \texttt{flatMap} and \texttt{map} transformations; \texttt{ExtractLinks} and \texttt{ExtractDomain} are UDFs).
A filter transformation is applied to discard empty output, and the results are then aggregated by count:\ \texttt{countItems} is another convenience transformation that AUT provides (essentially serving as syntactic sugar for a \texttt{groupBy} and \texttt{count}).
Another filter is applied to discard all target domains that receive five or fewer inlinks (to reduce the amount of noise in the webgraph) before the final (source domain, destination domain, count) output tuples are materialized and saved to a text file.
Here, we can see exactly how activities in our process model (Section~\ref{section:4activities}) translate into RDD transformations in an AUT script, although in this case the script does not generate a visualization, but rather stores the output for subsequent consumption elsewhere.

Given this script, the Spark engine orchestrates execution over collections at scale.
Although Spark was designed to run on clusters, we have primarily processed collections on individual multi-core servers.
This decision makes sense for a few reasons:\ since our project uses transient cloud virtual machines, spinning up and down clusters on demand adds an additional level of configuration complexity that is not strictly needed for our use case.
Spark is still able to make use of multiple cores on a single machine to analyze collections in parallel.
Most collections can be processed by individual servers within reasonable amounts of time, and furthermore our jobs are not latency sensitive.
For even our largest collections (see Section~\ref{section:cloud}), Spark has proven to be robust and has no problem with job scheduling or task management---it is simply a matter of time waiting for jobs to complete.
Our users are a patient lot.

\begin{figure}[t]
\vspace{0.2cm}
\centering
\includegraphics[width=0.95\columnwidth]{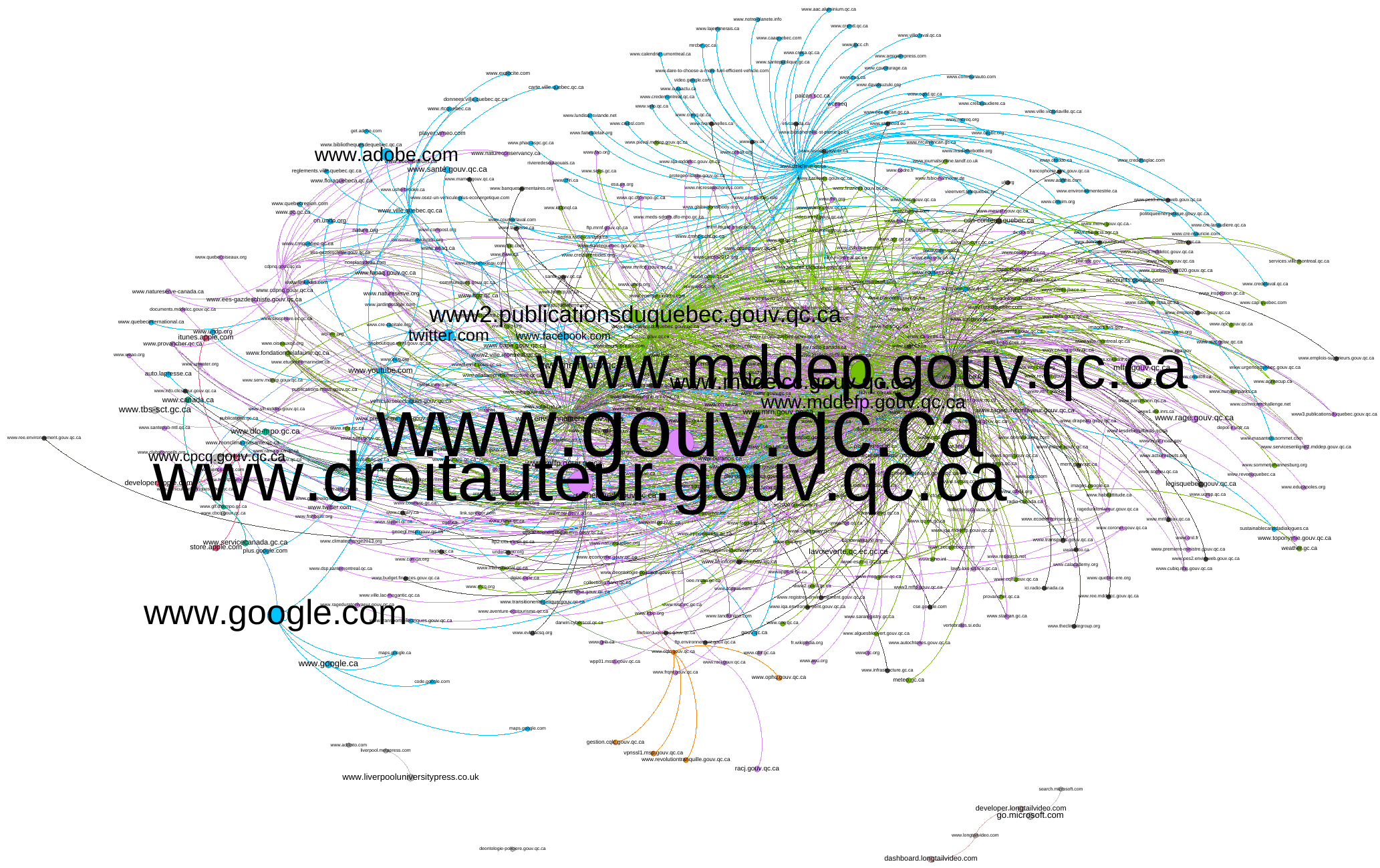}
\vspace{-0.2cm}
\caption{A visualization of the domain webgraph from a web archive of the Ministry of Environment of Qu\'ebec, collected between 2011 and 2014 by the Biblioth\'eque et Archives nationales du Qu\'ebec.}
\label{fig:gephi-network}
\vspace{-0.4cm}
\end{figure}

To complete this walkthrough, the domain webgraph can be ingested into the popular open-source network analysis platform Gephi~\cite{gephi} for visualization and further exploration.
The software has robust documentation and an active user base within the digital humanities and computational social sciences, making it a natural platform for scholars interested in web archives.
In this example, we use a web archive of the Ministry of Environment of Qu\'ebec, collected between 2011 and 2014 by the Biblioth\'eque et Archives nationales du Qu\'ebec.
After a series of basic transformations in Gephi---in this case, sizing the nodes and labels based on computed PageRank values---the scholar can see the basic outlines of the collection's hyperlink structure, shown in Figure~\ref{fig:gephi-network}.
Potential research questions begin to emerge:
For example, almost all hyperlinks are to pages within the \texttt{qc.ca} domain; there are few external ones beyond a few web services or platforms (Twitter for embedded accounts or Google).
Given the history of Canadian federalism, which has seen uneasy relationships between federal and provincial levels of government, it is notable that there are \emph{no} links to Environment Canada, their federal counterpart.
Already, a research project---looking at various provincial ministries to explore changing relationships with federal counterparts---begins to take shape.

Development efforts on the toolkit have followed standard best practices for open-source software.
Source code is held in a public GitHub repository,\footnote{\url{https://github.com/archivesunleashed/aut}} where we extensively use ``issues'' to keep track of bugs, feature requests, as well as for planning new features and discussing high-level design.
All issues are public and participation is open to everyone who may be interested---while (quite obviously) our team is the most active in the online forum, we regularly receive comments and feedback from the community.
Modifications to the codebase occur via pull requests and undergo code review before they are merged to the master branch.
The toolkit has fairly good unit test coverage, and standard continuous deployment tools simplify automated testing as part of the code review process.
We create official releases periodically using standard toolchains.

To conclude our discussion of the toolkit, we present two ongoing development efforts:

\smallskip \noindent {\it Transitioning from RDDs to DataFrames.}
While RDDs provide the core data structure in Spark, the platform has seen the emergence of DataFrames as an alternative higher-level abstraction for manipulating large collections of records.
The primary difference is that DataFrames conform to schemas and are organized into named columns, much like tables in a relational database, whereas RDDs can comprise heterogeneous records of arbitrary format.
While RDDs are more flexible---in that they support arbitrary record-level transformations---this flexibility is rarely needed by scholars, and in fact can lead to confusion and unexpected errors (for example, type mismatches).
With DataFrames, the scholar can refer to fields using meaningful names like the \texttt{src} and \texttt{dest} of a hyperlink, as opposed to RDDs, where they must use Scala's underscore notation (e.g., \texttt{r.\_1}) to access individual fields in a tuple.
Furthermore, Spark DataFrames were inspired by DataFrames in the highly-popular Pandas package for data analysis in Python.
Many scholars are already familiar with Pandas DataFrames, and thus they would be comfortable manipulating Spark DataFrames with minimal training.
This flattens the learning curve and lowers barriers to adoption.
There are additional performance benefits as well:\ conformance to schemas allows the Spark engine to safely make certain optimizations that would not be possible with RDDs, and thus certain operations with DataFrames may become quicker to execute.

We are currently in the process of replicating RDD features using DataFrames in Scala, with the goal of providing two different approaches to analyses; i.e., everything that can be done with RDDs will have a counterpart using DataFrames.
More concretely, this involves creating different DataFrame ``views'' on the raw archive records, since our input connectors are still written in terms of RDDs and we are leveraging Spark's internal machinery to build DataFrames from RDDs.
For example, the \texttt{webgraph} view presents a table of hyperlinks in a collection, with four columns:\ the crawl date, the source URL, the destination URL, and the anchor text.
Using this DataFrame, we can extract the domain webgraph with the simplified script shown on the right of Figure~\ref{fig:aut-example-code}.
The juxtaposition of the RDD version and the DataFrame version highlights some of the advantages of DataFrames:\  the scholar can refer to named columns using the dollar sign (\texttt{\$}) notation, which allows her to more easily track the course of each datum through the analytical flow.
Although conceptually, the records are undergoing similar transformations, the DataFrames code is simpler and easier to understand.
We have verified that execution of the DataFrames script is no slower than the RDD version, and thus clarity does not come at the cost of performance.

\smallskip \noindent {\it Transitioning from Scala to Python.}
As Spark was implemented in Scala, we followed suit and built AUT primarily in Scala as well.
Although scholars are less likely to be familiar with Scala (compared to Python or R), we have not found the language choice to be an insurmountable barrier based on experience from our datathons.
Nevertheless, it would be desirable to allow scholars to conduct analyses in Python, which is the language they are most likely to already know.
This is possible with PySpark, which provides a Python interface to Spark, and we are currently in the process of transitioning to Python as the default language when using the toolkit.
Specifically, we aim to replicate in Python all existing functionalities in Scala.
In conjunction with DataFrames, the next iteration of the toolkit will be more intuitive and familiar to scholars, further reducing the barriers to adoption.

\section{The Cloud}
\label{section:cloud}

The Archives Unleashed Cloud (AUK) was conceived as the ``canonical deployment'' of the Archives Unleashed Toolkit.
Although AUT is open source, we anticipated that installing, configuring, and deploying the toolkit might pose too high a barrier of entry for most scholars.
Instead, we aimed to create a ``cloud portal'' whereby scholars could easily ingest their collections and leverage the capabilities of AUT, while we handled configuration, maintenance, and deployment transparently behind the scenes.
Currently, AUK is built with tight integration with Internet Archive's Archive-It service:\
As discussed in Section~\ref{section:history}, this represents the greatest potential for achieving impact.

\begin{figure}[t]
\centering
\includegraphics[width=1.0\columnwidth]{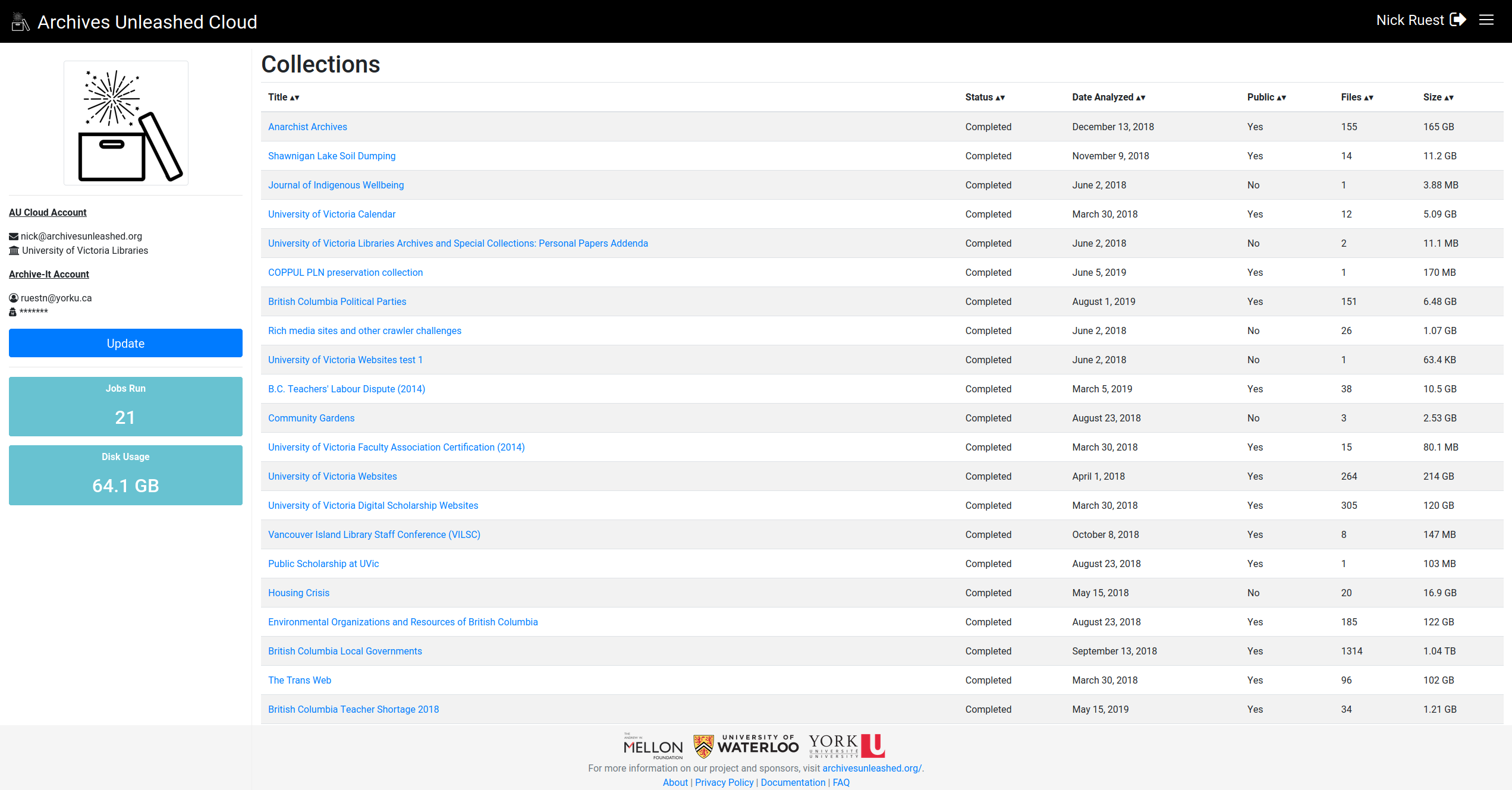}
\caption{A screenshot from the Archives Unleashed Cloud, showing the user dashboard that provides an overview of collections available for analysis.}
\label{fig:auk-dashboard}
\end{figure}

From the technical perspective, the Archives Unleashed Cloud (AUK) is a Rails application. 
Users can log in via their GitHub or Twitter credentials (the two current authentication methods we support), and are then presented with a basic dashboard interface.
Here, users provide their Archive-It credentials, which triggers a background job that imports metadata from the collections in their Archive-It account.
Once this job is finished, the user is notified via email that they can now analyze any of their Archive-It collections.
A screenshot of this dashboard interface is shown in Figure~\ref{fig:auk-dashboard}.

Once a collection is selected for analysis, AUK triggers a chain of background jobs to perform the actual computation.
These jobs physically execute on infrastructure provided by Compute Canada, which is a service that provides Canadian researchers with computing resources.
First, the raw WARC (or ARC) files comprising the collection are copied over to our Compute Canada storage via Archive-It's Web Archiving Systems API (WASAPI) data transfer endpoint.
Once the download job is finished, AUK coordinates the generation of the standard derivatives discussed in Section~\ref{section:derivatives} using the toolkit.
This is accomplished by provisioning a single-node server for Spark execution, as described in Section~\ref{section:toolkit}.

\begin{figure}[t]
\centering
\includegraphics[width=1.0\columnwidth]{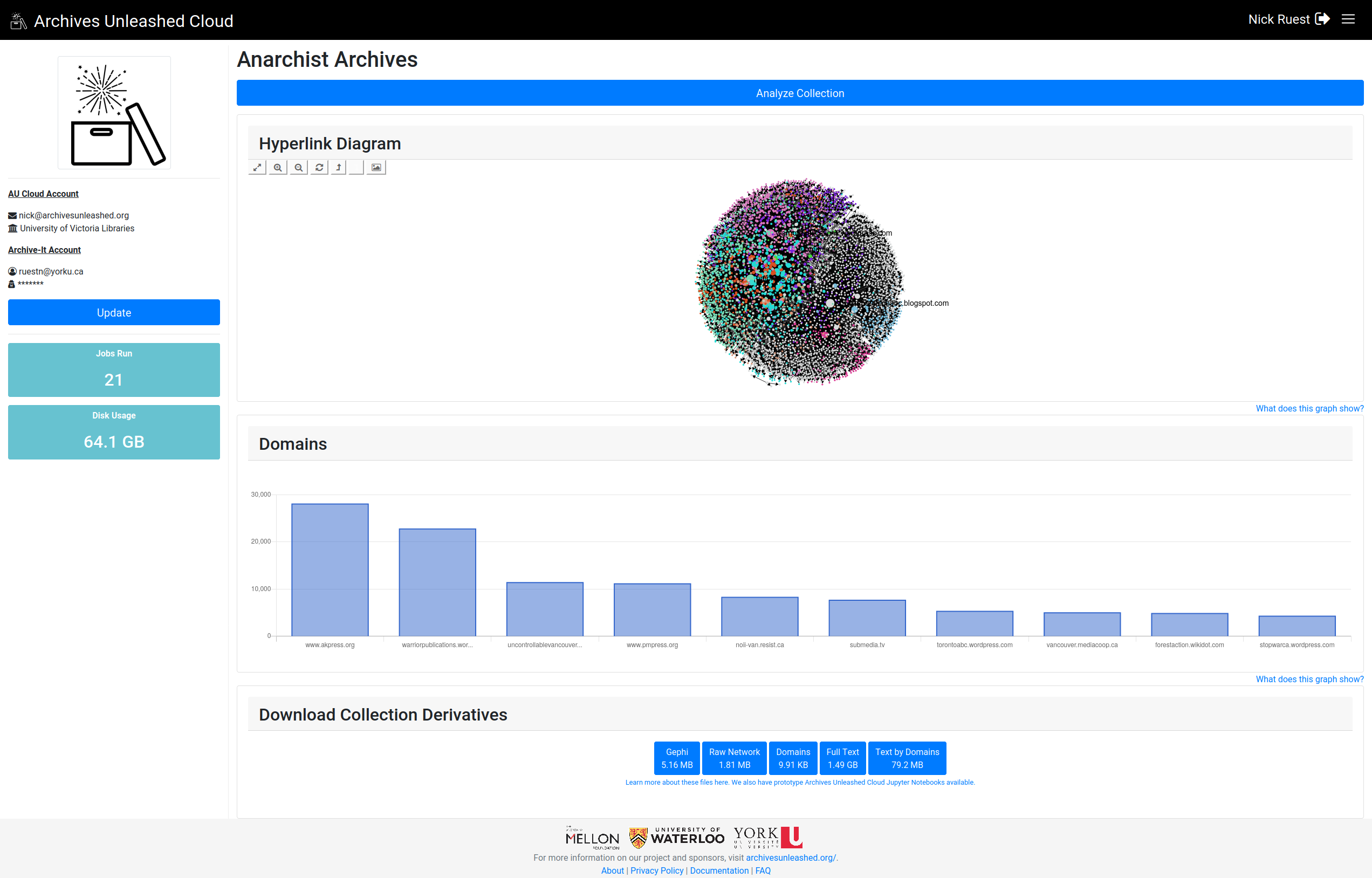}
\caption{A screenshot from the Archives Unleashed Cloud showing the overview of a collection, which provides a domain webgraph visualization, distribution of top domains, and links for downloading derivatives.}
\label{fig:auk-collection}
\vspace{-0.25cm}
\end{figure}

A few smaller jobs follow, and the user is notified via email once the entire processing pipeline has finished.
The final product is a collection overview page:\ an example for the Anarchist Archives collection from the University of Victoria is shown in Figure~\ref{fig:auk-collection}.
For collections that are manageable in size, we provide a JavaScript-based visualization of the domain webgraph;
a bar chart showing the top 10 domains in the collection is also provided.
Finally, the overview page provides download links for the derivative products that were created by the toolkit.
Currently, these products are available in CSV format, but we are in the process of adding support for Apache Parquet, a popular columnar storage format for big data.
The webgraphs are also available in a format that can be directly read by the Gephi graph visualization platform.

The Archives Unleashed Cloud itself is open source and publicly available on GitHub,\footnote{\url{https://github.com/archivesunleashed/auk}} and thus anyone could run their own private instance if desired.
The development of AUK follows the same best practices as the toolkit, already discussed in Section~\ref{section:toolkit}.

To conclude our description of the Archives Unleashed Cloud, we provide some statistics quantifying the extent of our efforts:\
Starting in August 2018, as of January 2020, the platform has processed approximately 1100 Archive-It collections from around 200 users.
These collections comprise 1.4 million individual files totaling 284 terabytes.
Analyses of these collections has translated into around 5600 individual jobs, representing a cumulative processing time of approximately 23,000 hours on Compute Canada servers.
The largest individual collection, the 2012 Summer Olympics Collection\footnote{\url{https://archive-it.org/collections/5713}} from the International Internet Preservation Consortium (IIPC), totals 17.6 TB and took around 900 hours (38 days) to process.
These statistics illustrate the robustness and scalability of the Archives Unleashed Toolkit and the Archive Unleashed Cloud as a portal to its capabilities.

\section{Notebooks}
\label{section:notebooks}

Summarizing the value proposition of the Archives Unleashed Toolkit and Cloud at a high level:\
We offer to an institutional subscriber of Archive-It the answer to the question, ``We've harvested a number of interesting web archives, now what?''
With a few keystrokes and a few clicks of the mouse, these collections can be distilled into useful derivatives that would serve as the starting point of scholarly inquiry.
These derivatives can be downloaded from the cloud portal, and in most cases they are manageable on scholars' laptops with tools of their choice.

How could we further lower barriers to entry?
There are two recent developments worth discussing:
First, the collecting institution can publish these derivatives on a data archiving site such as Zenodo or Dataverse.
See, for example, one such record on Zenodo for the Ministry of Environment of Qu\'{e}bec (2011--2014) Web Archive Collection Derivatives ~\cite{ruest_nick_2020_3599771}.
We are building such a publishing feature directly within AUK to simplify the workflow for content owners.
There are a couple of advantages for doing this:\ the datasets are now completely decoupled from AUK and available on the public web, thus increasing discoverability and broadening access.
Also, the dataset receives a citeable DOI and becomes a first-class citizen in the academic ecosystem:\ this incentivizes content creators because, for example, usage can now be tracked using standard bibliometric techniques.

The second major development we are excited about is connecting the derivative products directly to ``notebooks'' (Jupyter Notebooks being the most popular), which have emerged as a popular tool for data science.
Notebooks are typically provided on a web-based platform in an interface that interleaves code fragments (most often, in Python), descriptive prose (for example, describing a particular approach), and execution results (for example, graphs and figures).
Notebooks support rapid interactions with minimal setup; they can be saved, shared, and re-executed easily, supporting collaboration and reuse.

Building on our preliminary ``madlibs'' or fill-in-the-blank approach with notebooks~\cite{Deschamps2019}, we have working prototypes\footnote{\url{https://github.com/archivesunleashed/notebooks}} that take a collection's derivatives and automatically pre-populates a notebook with example analyses.
The notebooks can be downloaded and run locally or imported into a cloud service, offering the simplest implementation of the ``opening moves'' concept (see Section~\ref{section:4activities}) that we have so far devised.

\section{Related Work}

We are, of course, not the only ones who have tackled the challenge of providing scholarly access to web archives.
From the scholarly perspective, Br\"ugger has extensively theorized the use of the archived web as a research object, most notably in his recent monograph~\cite{Brugger:2018}; so too has Milligan (one of the co-authors), who conceptually explores methods for working with web archives~\cite{Milligan:2019}.
Weber and Napoli recently published a ``methodological approach to utilizing Web archives as a means of examining change in the news media industry'' in a paper focusing on local news~\cite{weber2018}.
While of interest to social scientists, they unfortunately do not provide a platform for other researchers to directly work with their tools. 
In the United Kingdom, the 2013--15 Big UK Domain Data for the Arts and Humanities (BUDDAH) project represented a pivotal engagement and collaboration between researchers and librarians interested in facilitating access to the archived web; the series of case study projects that came out of BUDDAH have informed considerable scholarship~\cite{winters2017} as well as the `Shine' search interface~\cite{jackson2016}.
Finally, as a book informed heavily by the BUDDAH case studies, the 2017 edited volume \emph{The Web as History:\ Using Web Archives to Understand the Past and the Present} provides a good starting point to explore various computational analyses of web archives~\cite{webashistory}.

From the technical perspective, the most notable related analytics platform is ArchiveSpark~\cite{holzmann2016}, also built on Apache Spark, which ``provides efficient access to Web archive data for extraction and derivation of smaller datasets''.
While our toolkit and ArchiveSpark share similar technical approaches, our emphasis on broader scholarly engagement is a principal difference. 
In addition, there is a wide array of utilities that can provide access to WARC files; these are documented on the collaboratively-maintained International Internet Preservation Consortium's ``Awesome List''.\footnote{\url{https://github.com/iipc/awesome-web-archiving}} 

\section{Looking Ahead}

What lies in store for the Archives Unleashed project moving forward?
We can currently identify two distinct classes of users:\ ``power users'' who use the toolkit directly to access its full capabilities (by extension, these are users who manage their own computational infrastructure), and other users who consume the derivatives with the toolkit or their tool of choice, possibly in a cloud notebook or on their own machines.
The second class of users is much larger and more diverse; for these scholars, we have eliminated barriers that are specific to web archives.
Thus, the road to adoption is smoothly paved.

However, what happens if a scholar who begins inquiry with the derivatives finds that they are no longer sufficient?
For example, her questions cannot be answered with the standard derivative products, and the information she is after must be extracted from the raw WARCs?
Or, she is working with a very large collection and the standard derivatives cannot be comfortably manipulated on her laptop?

Addressing the gap between these two current classes of users is one focus of our ongoing efforts.
We have two responses:
First, a scholar who finds the standard derivatives inadequate must have, by definition, already accumulated sufficient experience working with web archives.
We expect that technical adeptness will grow with the increasing sophistication of analysis.
At some point, learning how to install, maintain, and use the toolkit directly will become a worthwhile investment of effort, because by that time, the value of the software in enabling scholarly inquiry will already have been demonstrated.
Thus, it is not far-fetched to see a historian or social scientist ``take the plunge'' into becoming a ``power user''.

Second, the set of derivatives available for download from the Archives Unleashed Cloud needs not be static.
In response to growing community demand, we could make more derivatives available.
One concrete example we can point to involves image analysis.
In the early days of the project, most scholarly inquiry focused on HTML pages, both because they were a natural starting point and because the capabilities of the toolkit revolved mostly around text.
However, as the project matured, we cast our sights further and began to explore different aspects of image analysis (e.g., deduplicating images via checksums, finding the most popular image in a collection).
New toolkit capabilities stimulated further scholarly demand---such that we are currently working through the implications of providing image links as a standard derivative.
Of course, each additional derivative involves incrementally more processing and storage, and so a balance between resource consumption and scholarly demand must be struck.
However, our broader point is that what's provided and what's desired can evolve in a mutually informing way.
Once again, the analogy of opening moves in chess is perhaps helpful.
The derivatives available in the Archives Unleashed Cloud may be viewed as a compilation of the most popular opening moves, but as scholars begin ``playing new openings'', our platform can evolve to broaden its repertoire.

To conclude, we have come to realize over the course of the project that our goal of broadening scholarly access to web archives is not a destination we can arrive at, but rather an ongoing process.
The task of expanding accessibility is likely never ending:\ as a simple example, familiarity with Python and other programming languages exists along a wide spectrum, and there are always less technologically-prepared scholars we'd like to reach.
We believe that our project has made substantial progress, but there is still a long road ahead.

\section{Acknowledgments}

This research was supported by the Andrew W.\ Mellon Foundation, the Social Sciences and Humanities Research Council of Canada, as well as Start Smart Labs, Compute Canada, the University of Waterloo, and York University.
We'd like to thank Jeremy Wiebe, Ryan Deschamps, and Gursimran Singh for their contributions.



\end{document}